\documentclass[referee]{aa}  
\usepackage{amsmath,amsfonts,amssymb}

\usepackage{graphicx}

\usepackage[british]{babel}

\usepackage{natbib}

\usepackage{txfonts}

\begin{document}

\title{Gravitational waves and lensing of the metric theory proposed by Sobouti}

\titlerunning{Waves and lensing in Sobouti's \( f(R) \) metric theory}

\author{ S. Mendoza
          \and
          Y.M. Rosas--Guevara 
          }

\offprints{S. Mendoza}

\institute{Instituto de Astronom\'{\i}a, Universidad Nacional 
       Aut\'onoma de M\'exico, AP 70-264, Distrito Federal 04510,
       M\'exico \\
\email{sergio@astroscu.unam.mx, yetli@astroscu.unam.mx}
}

\date{\today}

\abstract
{}
{ We investigate in detail two physical properties of the metric \( f(R)
\) theory developed by Sobouti (2007).  We first look for the possibility of 
producing gravitational waves that travel at the speed of light.  We then
check the possibility of producing extra bending in the lenses produced by
the theory.
}
{  We do this by using standard weak field approximations to the
gravitational field equations that appear in Sobouti's theory.
}
{ We show in this article that the metric theory of gravitation
proposed by Sobouti (2007) predicts the existence
of gravitational waves travelling at the speed of light in vacuum.
In fact, this is proved in general terms for all metric theories of
gravity which can be expressed as powers of Ricci's scalar.
We also show that an extra additional lensing as compared to the one
predicted by standard general relativity is produced.
}
{
These two points
are generally considered to be of crucial importance in the development
of relativistic theories of gravity that could provide an alternative
description to the dark matter paradigm.
}

\keywords{ Gravitation -- Gravitational waves -- Gravitational lensing -- 
Relativity}

\maketitle
%

\section{Introduction}
\label{introduction}

  One of the greatest challenges of modern astrophysics is the validation
of the dark matter paradigm \citep{bertone04}.  Postulating the
existence of non--barionic dark matter has given a lot of success to
many astrophysical theories.  However,  no matter how hard the strange
matter has been looked for, it has never been directly observed, nor
detected \citep{munoz,cooley06}.

  It was \citet{milgrom83} who proposed that, to understand certain
astrophysical observations it was necessary to change Newton's law
of gravitation.  With time, this idea has developed strongly up to the
point of building a relativistic Tensor-Vector-Scalar (TeVeS) theory
that generalises and substantiates the Modified Newtonian Dynamics (MOND)
ideas introduced by \citeauthor{milgrom83} \citep{bekenstein04}.

  The complications introduced by TeVeS have lead different groups
to think of an alternative possibility.  This has been motivated
by recent development on metric theories of gravity applied to the
problem of dark energy.  Some researchers \citep[cf.][and references
therein]{capozziello02,capozziello03,nojiri03,carroll04,capozziello05,capozziello06,nojiri06}
have shown that it is possible to explain different cosmological
observations without the need of dark energy.  The idea is to introduce
a general function \( f(R) \) in the Einstein--Hilbert action, instead
of the standard Ricci scalar \( R \).  The resulting differential
equations that appear due to this introduction are of the fourth order.
This introduces a degree of complexity on the field equations, but makes
it possible to reproduce some results that are usually thought of as
being due to a mysterious dark energy field.

  In the same sense, \citet{capozziello06grav} and \citet{sobouti06} have
developed two different \( f(R) \) theories that can reproduce the
anomalous rotation curves produced in different spiral galaxies.
The advantages of \citeauthor{sobouti06}'s description are many.
His theory reproduces naturally the standard Tully--Fisher relation,
it converges to a version of MOND and, due to the way the theory is
developed, the resulting differential equations are of the second order.
More importantly, \citet{capozziello06grav} showed that \( f(R) = R^n \),
with \( 1.34 \lesssim n \lesssim 2.41 \) reproduces rotation curves of a
number of spiral galaxies.  On the other hand, \citet{sobouti06} showed
that if \( f(R) = R^{ \left( 1 - \alpha / 2 \right) } \), then different
rotation curves associated to spiral galaxies can be accounted for.
Also, since \( \alpha \ll 1 \), the modification can be thought of as a
small deviation to the Einstein--Hilbert action, i.e.  \( f(R) \approx
R \left[ 1 - \left( \alpha / 2 \right) \ln R + \left( \alpha / 2 \right)
\ln\left( 3 \alpha \right) \right] \).

  Central to the development of a good modified theory of gravity that
can describe the phenomenology usually ascribed to dark matter, is the
analysis of the propagation of gravitational waves and the amount of
lensing implied by the theory.  In this article, we show that all \(
f(R) = R^n \) metric theories of gravity produce gravitational waves
that propagate at the velocity of light \( c \) in vacuum.  This is
a crucial step in order to consider Sobouti's theory as a possible
alternative to the dark matter problem and so, it can begin to be applied
to real astrophysical situations.  Relativistic theories of MOND have
been proposed in the past (see e.g. \citet{bekenstein06} and references
therein) that show superluminal propagation of the waves produced by
the fields and so, they were rejected immediately.  For example, one
of the crucial steps in using TeVeS as an alternative to dark matter is
that the waves produced by the theory are never superluminal.  We also
show that the theory proposed by Sobouti describes additional
lensing to the standard general relativistic version.  In other words,
the theory developed by Sobouti can in principle be considered
an alternative theory, in order to do astrophysical comparisons with
current models of dark matter.   Also, this theory may challenge what
TeVeS has been trying to explain in certain astrophysical situations
and so, astrophysical predictions between TeVeS and Sobouti's theory
must be done in the future.

\section{Modified field equations}

  The alternative gravitational model used in what follows is one that
introduces a modification in the Einstein--Hilbert action as follows (see
e.g. \citet{sobouti06} and references therein):

\begin{equation}
  S = \int \left\{ \frac{ 1 }{ 2 } f(R) + L_\text{m} \right\} \, \sqrt{-g}
    \, \mathrm{d}^4 x, 
\label{eq001}
\end{equation}

\noindent where \( L_\text{m} \) is the Lagrangian density of matter
and \( f(R) \) is an unknown function of the Ricci scalar \( R \).
Variation of the action \( S \) with respect to the metric \( g_{\mu\nu}
\) gives the following field equations \citep{capozziello03}

\begin{equation}
  R_{\mu\nu} - \frac{ 1 }{ 2 } g_{\mu\nu} \frac{ f }{ F } = \left(
    F_{;\mu\nu} - F_{;_\lambda}^\lambda g_{\mu\nu} - T_{\mu\nu} \right)
    \frac{ 1 }{ F },
\label{field-equations}
\end{equation}

\noindent where \( F := \mathrm{d} f / \mathrm{d} R \) and \( T_{\mu\nu}
\) is the stress--energy tensor associated to the Lagrangian density of
matter \( L_\text{m} \) and \( R_{\mu\nu} \) is the Ricci tensor.

  In order to apply to galactic systems, the metric is chosen as a
Schwarzschild--like one given by \citep{cognola06,sobouti06}

\begin{equation}
  \mathrm{d} s^2 = -B(r) \mathrm{d} t^2 + A(r) \mathrm{d} r^2 + r^2 \left(
    \mathrm{d} \theta^2 + \sin^2 \theta \, \mathrm{d} \varphi^2 \right).
\label{eq15}
\end{equation}

  \citeauthor{sobouti06} showed that the combination \( A(r) B(r) = g(r) \)
and so \( g(r) = ( r / s )^\alpha \approx 1 + \alpha \ln\left( r / s
\right) \), with the Schwarzschild radius \( s := 2 G M \).  His
calculations also show that  the functions \( A(r) \) and \( B(r) \)
applied to galactic phenomena are given by

\begin{gather}
   \frac{ 1 }{ A } = \frac{ 1 }{ \left( 1 - \alpha \right) } \left[ 1 -
   \left( \frac{ s }{ r } \right)^{ \left( 1 - \alpha / 2 \right)}
       \right],
	       					\label{eq-for-A} \\ 
   B = \left( \frac{ r }{ s } \right)^\alpha \frac{ 1 }{ A },
\label{eq-for-B} 
\end{gather} 

  For the sake of simplicity we think of \( f(R) \) as
given by \citeauthor{sobouti06}'s~(\citeyear{sobouti06}) model (but see
\citet{nojiri04} for the first model that introduces a logarithmic \( f(R)
\) in cosmology), i.e., 

\begin{equation}
  f(R) = R^{ \left( 1 - \alpha / 2 \right) } \approx R \left[ 1 - \frac{
    \alpha }{ 2 } \ln R + \frac{ \alpha }{ 2 } \ln\left( 3 \alpha \right)
    \right].
\label{eq005}
\end{equation}
 
\noindent The parameter \( \alpha \) is chosen in such a way that \(
F(r,\alpha) \rightarrow 1 \), which corresponds to standard general
relativity,  as \( \alpha \rightarrow 0 \).

\section{Gravitational waves}
\label{gravitational-waves}

  Just as it happens in standard general relativity, it is expected
that a modified metric theory of gravity predicts gravitational waves.
These should propagate through space--time with velocity equal to that
of light.  We now show that this happens for all cases in which \( f(R) =
R^n \), where \(n\) is any number.  To do so, we consider a space--time
manifold with a metric \( g_{\mu\nu} \) deviating by a small amount  \(
h_{\mu\nu} \) from the Minkowski metric \(\eta_{\mu\nu}\) in such a
way that

\begin{equation}
   g_{\mu\nu}=\eta _{\mu\nu}+h_{\mu\nu},  
\label{eq1}
\end{equation}

\noindent with  \(\mid h_{\mu\nu}\mid \ll 1\). Using this we can make
arbitrary transformations of the coordinates \( x^\mu \) (or reference
frame) in such a way that \( x^{\mu} \rightarrow x^{\mu} + \xi^{\mu}
\), with \( \xi^{\mu} \) small.  As usual, we impose the Lorentz
gauge condition:

\begin{equation}
  \Psi^{\mu\nu} \ _{,\mu}=0,
\label{eq2}
\end{equation} 

\noindent with \( \Psi^{\mu\nu} := h^{\mu\nu} - (1/2) \, h \, \eta^{\mu\nu}
\) and \(h:=h^{\mu }_{\mu }\). The Ricci tensor \( R_{\mu\nu} \) and the
Ricci scalar \( R \) to first order in \( h_{\mu\nu} \) are consequently
given by

\begin{equation}
  R_{\beta\nu}\approx \frac{1}{2}\big[ \Psi _{\alpha \nu , \beta }\ 
    ^{\alpha}+\Psi _{\beta \mu ,} \ ^{\mu } \ _{\nu }-\Psi _
    {\beta \nu ,\alpha }\ ^{\alpha }
    +\frac{1}{2}\eta _{\beta\nu }\Psi _{,\alpha } \ ^{\alpha }\big ],
\label{eq3}
\end{equation} 

\begin{equation}
   R\approx \frac{1}{2}\big[2 \Psi _{\alpha \nu , }\ ^{\alpha \nu }
   +\Psi _{,\alpha }\ ^{\alpha }\big ].
\label{eq4}
\end{equation}

\noindent With the substitution on equations \eqref{eq3} and \eqref{eq4}
in the field equations \eqref{field-equations} with \( f(R) = R^n \)
we obtain

\begin{equation}
  \begin{split}
   ( \Psi_{\beta\nu} )_{,\, \alpha} \! ^{\alpha} = & 
     -\frac{1}{6}\eta _{\beta\nu}\Bigg (\frac{1+n}{n} \Bigg ) 
     ( \Psi _{,\,\alpha }\! ^{\alpha } ) + (n-1)(\Psi _{,\,\alpha
     }\! ^{\alpha })^{-1} (\Psi _{,\, \alpha }\! ^{\alpha })_{,\, \beta\nu} 
     + \\ 
  & +(n-1)(n-2)(\Psi _{,\,\alpha }\! ^{\alpha })^{-2}
    ( \Psi _{,\, \alpha }\! ^{\alpha } )_{,\,\beta} (\Psi _{,\,\alpha
    }\! ^{\alpha })_{,\,\nu}.
  \end{split}
\label{eq5}
\end{equation} 

  Because of the fact that  \( \Psi = \eta^{ \kappa \mu } \Psi_{
\kappa \mu } \), the gauge condition \eqref{eq2} and the commutativity
of partial derivatives to first order in \(h_{\mu\nu}\), the right hand side
of equation~\eqref{eq5} is null. Therefore we obtain the ordinary  wave 
equation, i.e. \( \Psi_{\beta\nu,\alpha}{}{}{}{}^{,\alpha} = 0 \).

 We now consider a weak perturbation relative to an arbitrary
metric  \( ^{(0)}\! g_{\mu\nu}\). Then, the metric \( g_{\mu\nu} \) takes 
the form

\begin{equation}
   g_{\mu\nu}\approx ^{(0)}\!\!\! g_{\mu\nu} + h_{\mu\nu},
\label{eq6}
\end{equation}

\noindent In this expression,  \( \mid h_{\mu\nu} \mid \, \approx
\text{O}(\lambda /L)\) where the wavelength \(\lambda\) is small compared
to an arbitrary characteristic length \( L \), which is related to curvature
of space--time.

  Let us define the tensor \(\Psi_{\mu\nu}:= h_{\mu\nu}-
^{(0)}\! h \, g_{\mu\nu}\) .  We impose the same transverse traceless gauge
condition \citep{daufields} as we did for equation~\eqref{eq2},
by replacing the standard derivative with a covariant one, i.e.

\begin{equation}
  \Psi^{\nu\mu}\ _{;\nu} = 0.
\label{eq7}
\end{equation}

  The corrections to the Ricci tensor and the Ricci scalar  to first order in
\(h_{\mu\nu}\) are  respectively given by

\begin{equation}
  R_{\beta\nu}\approx ^{(0)}\!\! R _{\beta\nu}+\Psi_{\beta\nu} \! 
   _{\, ;\,\alpha}  \ \! ^{\alpha}-^{(0)}\!\!\! 
   R_{\alpha\beta\mu\nu}h^{\alpha\mu},
\label{eq8}				
\end{equation} 

\begin{equation}
  R\approx^{(0)}\! R+^{(0)}\!\!\! R_{\mu\alpha}\Psi^{\mu\alpha}.
\label{eq9}
\end{equation} 

\noindent The terms \(^{(0)}\! R _{\beta\nu}\),
\(^{(0)}\! R_{\alpha\beta\mu\nu}\) and  \(^{(0)}\! R\) are calculated
with respect to the metric \( ^{(0)}\! g_{\mu\nu}\).  Substitution of
equations~\eqref{eq8} and \eqref{eq9} in the field equations with the
function \( f(R) = R^n \), we find to first order of approximation the
following relation:

\begin{equation}
   \begin{split}
       \frac{1}{2}\Psi_{\beta\nu\, ; \, \alpha}\ ^{\alpha} -
       & ^{(0)}\! \! R_{\alpha\beta\mu\nu}\Psi ^{\alpha\mu}- 
         ^{(0)}\!\! R \Psi _{\beta\nu }\approx - C_{1}
         \big ( \ ^{(0)}\! g_{\beta\nu} \ ^{(0)}\!\! R_{\mu\alpha}
         \Psi^{\mu\alpha}\big )+\\
      + & C_{2}\bigg \{ (^{(0)}\! R)^{-1}
         \big[\  ^{(0)}\!\! R_{\mu\alpha}\Psi^{\mu\alpha}\big ]
         _{\, ; \, \beta\nu}-(^{(0)}\! R)^{-2} \big[\ ^{(0)}\!\! 
         R_{\mu\alpha}\Psi^{\mu\alpha}\big ] 
         \big[\ ^{(0)}\! R\big ]_{\, ; \, \beta\nu} \bigg \} - \\
      -&C_{3}\bigg \{ (^{(0)}\!\! R)^{-2} \big[\ ^{(0)}\!\! R\big ] 
       _{\, ; \, \nu} \big[\ ^{(0)}\!\! R_{\mu\alpha}
       \Psi^{\mu\alpha}\big ] _{\, ; \, \beta}
       +(^{(0)}\!\! R)^{-2} \big[\ ^{(0)}\!\! R\big ] 
       _{\, ; \, \beta} \big[\ ^{(0)}\!\! R_{\mu\alpha}
       \Psi^{\mu\alpha}\big ] _{\, ; \, \nu}-\\
     - & 2 (^{(0)}\!\! R_{\mu\alpha}\Psi^{\mu\alpha})
       (^{(0)}\!\! R)^{-3}\big[\ ^{(0)}\!\! R\big ] _{\, ; \, \beta}
        \big[\ ^{(0)}\!\! R\big ] _{\, ; \, \nu} \bigg \} ,
  \end{split}
\label{eq10}
\end{equation}

\noindent where \(C_1\), \(C_2\) and \(C_3\) are constants for a fixed
value of \( n \). The terms involving  \(^{(0)}\! R _{\beta\nu}\),
 \(^{(0)}\! R_{\alpha\beta\mu\nu}\) and 
\(^{(0)}\! R\) as well as the ones involving first partial derivatives in
\(\Psi\) can be neglected due to the fact that \(\lambda /L \ll 1\). This
result becomes clearer if we propose a solution of equation
\eqref{eq10} as   

\begin{equation}
  \Psi_{\beta\nu}=\text{Re} \left( A_{\beta\nu} e^{ ( i\phi ) }\right), 
\label{eq11}
\end{equation} 

\noindent where the function \(\phi\) is the eikonal. For the particular
case we are dealing with, the eikonal
is large if we are to satisfy the condition \( \lambda / L \ll 1 \). 

  We now define the 4--wavevector \( k_\mu \) as

\begin{equation}
  k_{\alpha}:=\phi_{;\alpha}.
\label{eq12}
\end{equation}

\noindent Substitution of equation~\eqref{eq11} in  \eqref{eq10} and
keeping the dominant terms to second order in \(\phi\) we find the 
following relation:

\begin{equation}
   k_{\alpha}k^{\alpha}=0,
\label{eq13}
\end{equation}

\noindent i.e. the 4--wavevector is null.  Therefore,
gravitational waves in a \( f(R) = R^n \) metric theory of gravity
propagate at the speed of light \( c \).

\section{Gravitational lensing}
\label{lensing}

  A key part of gravitational lensing is the light bending angle due to
the gravitational field of a point-like mass. This can be determined
by using the fact that light rays move along null geodesics, i.e. \(
\mathrm{d}s^2 = 0\).  Similarly, its trajectory must satisfy:

\begin{equation}
  g_{\mu\nu}\frac{{\rm d} x^{\mu}}{{\rm d} \xi}\frac{{\rm d}
   x^{\nu}}{\rm d \xi}=0,
\label{eq14}
\end{equation}

\noindent where \(\xi\) is a parameter along the light ray.  We now
derive an expression for the bending angle by a static spherically
symmetric body.  The metric of the corresponding space--time is given
by the Schwarzschild--like metric \eqref{eq15}. Because of
the spherical symmetry, the geodesics of~\eqref{eq14} lie in a plane,
say the equatorial plane \(\theta=\pi /2\). The fact that the metric
coefficients do not depend neither on \(\varphi\) nor t, yields the
following equations:

\begin{equation} 
  r^2 \frac{\rm d \varphi}{\rm d \xi}=J, \qquad
  \frac{\rm d t}{\rm d \xi}= \frac{1}{B(r)}
\label{eq16}    
\end{equation}

\noindent where \(J\) is a constant of integration. Substitution of
equations~\eqref{eq15} and \eqref{eq16} into \eqref{eq14}  and replacing
\(\xi \) by \(\varphi\) as an independent variable, we obtain

\begin{equation}
  \frac{ 1 }{ B(r) } - \frac{ J^{2} }{ r^4 } A(r) \left( \frac{ \mathrm{d}
    r }{ \mathrm{d} \varphi } \right)^{2} - \frac{ J^2 }{ r^2 }=0,
\label{eq17}
\end{equation}

  If the closest approach to the lens occurs at a distance \( r_m \)
with an angle \(\varphi_m\), such that \( ( \textrm{d}r_m / \textrm{d}
\varphi ) = 0 \), then \( J \) is given by

\begin{equation} 
  J = \frac{ r_m }{ \sqrt{ B( r_m ) } }.
\label{eq18}
\end{equation}

\noindent  Equations \eqref{eq16} and \eqref{eq17} then yield 

\begin{equation} 
  {\rm d } \varphi = \frac{ { (AB ) }^{1/2}{\rm d }r}{r\sqrt{(r/r_{m})^{2}
     B(r_{m})-B(r)}}.   
\label{eq19}        
\end{equation}  

  We now consider a light ray that originates in the asymptotically
flat region of space--time and is deflected by a body before arriving
at an observer in the flat region. Therefore, equation \eqref{eq19}
yields the following expression for the bending angle \( \beta \):

\begin{equation} 
  \beta=2 \int ^{\infty}_{r_m} \frac{(AB)^{1/2} \mathrm{d} r}{r\sqrt{
    (r/r_{m})^{2}B(r_{m})-B(r)}}-\pi.
\label{eq20} 
\end{equation} 

  To compute the light bending angle in the \citet{sobouti06} \( f(R)
\) gravitation, we substitute the metric functions  \eqref{eq-for-A} and
\eqref{eq-for-B} into expression \eqref{eq20}.  If we now
define \( x := r_m / r \), the deflection angle takes the following form:

\begin{equation} 
   \beta =2 \int^{1}_{0}\frac{(sx/r_{m})^{-(\alpha/2)}{\rm d}
     x}{\sqrt{B(1)-x^{2}B(x)}}-\pi .
\label{eq23}
\end{equation} 

  This integral can be solved exactly if we assume that \( s / r_m \ll 1
\), in order to obtain:

\begin{equation}
 \begin{split}
  \beta & = 2 \sqrt{ 1 - \alpha } \, \left\{ \int_0^1 \frac{ \mathrm{d}
    x }{ x^{ \alpha/2 } \sqrt{ 1 - x^{ 2 - \alpha } } } + \frac{ 1 }{
    2 } \left( \frac{ s }{ r_m } \right)^{ 1 - \alpha/2 } \int_0^1 \frac{
    \left( 1 - x^{3 - 3 \alpha / 2 } \right)  \, x^{-\alpha/2} }{ \left( 1
    - x^{ 2 - \alpha } \right)^{ 3/2 } }  \, \mathrm{d} x \right\} - \pi,
    						\\
    & =  \pi \left[  \frac{ 2 \sqrt{ 1 - \alpha } }{ 2 - \alpha } - 1
        \right] + 2 \sqrt{ 1 - \alpha } \left( \frac{ s }{ r_m } \right)^{
	1 - \alpha / 2 }.
 \end{split}
\label{eq24}
\end{equation}

\noindent For \(\alpha=0\), this expression converges to the bending angle
expected in traditional general relativity. 

\begin{figure}
 \begin{center}
   \includegraphics[scale=0.7]{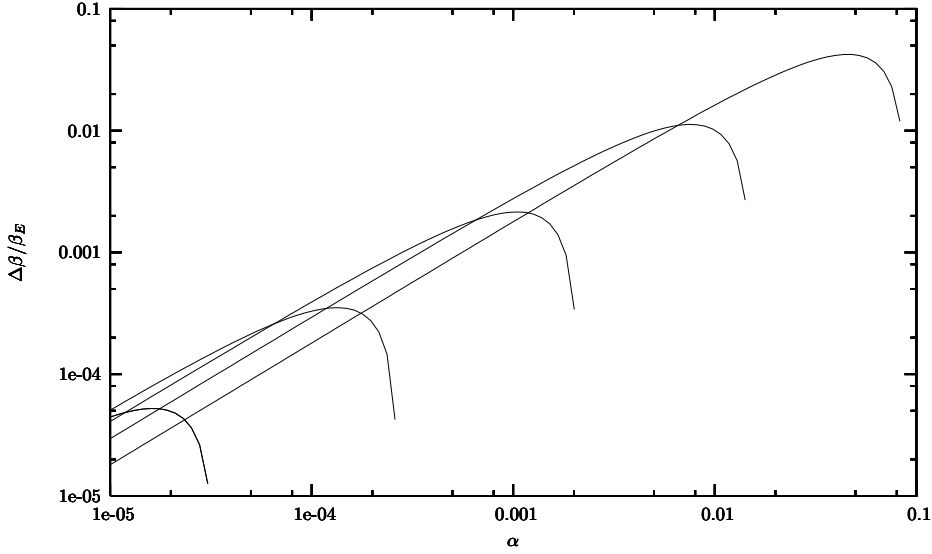}
 \end{center}
 \caption[Fluctuation of the bending angle]{The plot shows 
 fluctuations \( \Delta \beta / \beta_\text{E} \) for the bending angle 
 \( \beta \) as compared to \( \beta_\text{E} \), the one predicted by 
 general relativity.  From left to
 right, each plot corresponds to values of \( s / r_m \) given by \(
 10^{-6},\ 10^{-5},\ 10^{-4},\ 10^{-3},\ 10^{-2} \), respectively.  For
 greater values of \( s / r_m \) it is expected that more bending will be
 produced.}  
\label{fig0}
\end{figure}

  Figure~\ref{fig0} shows the fluctuation \( \Delta \beta \) given by

\begin{equation}
  \frac{ \Delta \beta }{ \beta_\text{E} } = \frac{ \beta - 
    \beta_\text{E} }{ \beta_\text{E} },
\label{eq25}
\end{equation}

\noindent where \( \beta_\text{E} \) is the bending angle obtained by
general relativity.   This fluctuation is a function of the parameter \(
\alpha \).  In fact, to \( \text{O}(\alpha) \) it follows that

\begin{equation}
  \frac{ \Delta \beta }{ \beta_\text{E} } = \frac{ 1 }{ 2 } \alpha \left\{
    \ln \frac{ r_m }{ s } - 1 \right\}.
\label{eq25a}
\end{equation}

  The Figure shows that in order to obtain significant bending, the
parameter \( s / r_m \) needs to be not too small for an appropriate value
of \( \alpha \).  Of course, when \( s / r_m \) is closer to \( 1 \),
more bending is expected.  The result obtained in equation~\eqref{eq24}
for the bending angle is only valid for \( s / r_m \ll 1 \) and so, the
plot cannot be used to test greater values of \( s / r_m \). However,
the trend seen in going from \( s / r_m = 10^{-6} \) to \( 10^{-2} \)
is strongly suggestive of significant \( \Delta \beta / \beta_\text{E}
\)  for yet smaller impact parameters.  This could in principle account
for anomalous lensing in clusters of galaxies considering that the
relevant \( \alpha \) at those scales might differ from the galactic
value calculated by \citeauthor{sobouti06}.  In fact, \citet{sobouti06}
showed empirically that

\begin{equation}
  \alpha = \alpha_0 \left( \frac{ M }{ M_\odot } \right)^{1/2},
\label{eq25b}
\end{equation}

\noindent with \( \alpha_0 \approx 2.8 \, \times \,  10^{-12} \).  Even if
one assumes that \( \alpha_0 \)  is a universal constant, then for 
very massive bodies it may be possible to obtain the required extra bending.

\section{Lensing framework}
\label{lensing-framework}

\begin{figure}
  \begin{center}
    \includegraphics[scale=0.45]{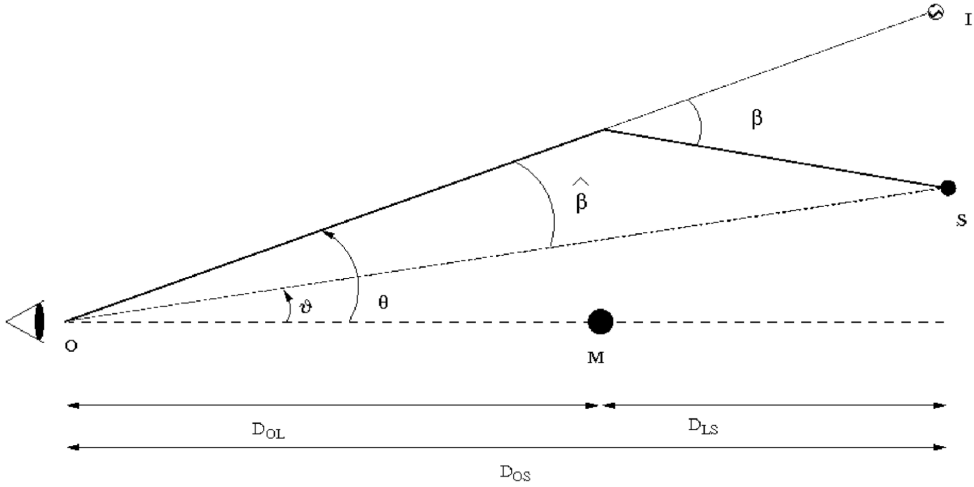}
  \end{center}
  \caption[Diagram of the lensing geometry]{Diagram of the lensing 
          geometry. \(\vartheta\) is the angular position of the
          source; \(\theta\) is the angular position of an image;
          and \(D_{OL}\),\(D_{OS}\) and \(D_{LS}\) are the observe-lens,
          observer-source and lens-source distances, respectively}.
\label{fig1}
\end{figure}

  To study how a body acts as a gravitational lens, we begin with a
through analysis of lensing  by a static, spherically symmetric body
with mass \(M\). Additionally, we assume the gravitational field
produced by the lens to be weak. Figure~\ref{fig1} gives a diagram of
the lensing situation and defines standard quantities: \(\vartheta\) is the
angular position of the source, \(\theta\) is the angular position of an
image and \( D_{OL} \), \( D_{OS} \) together with \( D_{LS} \) are the 
observer--to--lens, observer--to--source and lens--to--source distances
respectively.

  From the figure, elementary trigonometry establishes the lens equation:

\begin{equation} 
    \vartheta=\theta - \hat \beta, 
\label{eq27}
\end{equation} 

\noindent where \( \hat{\beta}  := \beta D_{LS} / D_{OS} \). 

  Since \( r_m=\theta D_{OL}\) and using equation~\eqref{eq27}, it then 
follows that the lens equation takes the form

\begin{equation}
  \Theta^2 - \Theta \left[ \varPhi + C_{1} \right] - C_2 \Theta^{ \alpha /
    2 } = 0.
\label{eq28}
\end{equation}

\noindent where 

\begin{displaymath}
  C_{1} := \frac{ \pi }{ \theta_E } \,  \frac{ D_{SL} }{ D_{OS} } \left[
    \frac{ 2 \sqrt{ 1 - \alpha } }{ 2 - \alpha } - 1 \right], \qquad
  C_2 := \theta_E^{ -\alpha / 2 } \left( \frac{ 1 }{ 2 } \frac{ D_{OS}
  }{ D_{LS} } \right)^{ -\alpha / 2 } \, \sqrt{ 1 - \alpha },
\end{displaymath}

\noindent and 

\begin{equation}
  \theta _{E} := \sqrt{  \frac{ 4 G M \, D_{LS} }{ D_{OS} \, D_{OL} } },
  \qquad
   \Theta := \frac{\theta}{\theta_{E}}, 
   \qquad 
   \varPhi := \frac{\vartheta}{\theta_{E}}.
\end{equation}

\noindent The quantities \(\Theta\) and \(\varPhi\) can be thought of as
``scaled angles'' with respect to \( \theta_E \).

  In order to solve the lens equation note that, to first order of
approximation, the solution can be written as

\begin{equation}
  \Theta=\Theta_{0}+\Theta_{1},
\label{eq29}
\end{equation}

\noindent where \(\Theta_{0}\) represents the standard image position
and \(\Theta_{1}\) is the correction term to first order. Substitution of
equation~\eqref{eq29} on \eqref{eq28} gives:

\begin{equation}
  \Theta_1= \frac{ C_1 \Theta_0 + C_2 \Theta_0^{ \alpha / 2 } - 1 }{ 
    2 \Theta_0 - \varPhi - C_1 - \left( \alpha /  2 \right) C_2  \, 
    \Theta_0^{ \alpha / 2 - 1 } }.
\label{eq30}  
\end{equation}.
    
 We can now obtain the magnification \( \mu := \theta \mathrm{d} \theta /
\vartheta \mathrm{d} \vartheta \) of a lensed image at angular
position \(\theta\).   Using equations~\eqref{eq29} and \eqref{eq30} it
follows that this magnification is given by

\begin{equation} 
  \mu = \frac{ \Theta_0 }{ \Phi } \frac{ \mathrm{d} \Theta_0 }{ \mathrm{d}
    \Phi } + \frac{ \Theta_1 }{ \Phi } \frac{ \mathrm{d} \Theta_0 }{
    \mathrm{d} \Phi }.
\label{eq31}
\end{equation}

  With the known positions and magnifications of the images it is now
straightforward to obtain the time delay \( \bigtriangleup \tau \) of a
light signal. This time is defined as the difference between the light
travel time \( \tau \) for an actual light ray and the travelled time \(
\tau_\text{eu} \) for an unlensed one, i.e.

\begin{equation} 
  \delta \tau = \tau - \tau_\text{eu}.
\label{eq32}
\end{equation}

  To compute the time delay, we use the weak field approximation
for the metric, that is \( g_{00} \approx 1 + 2 \phi \).  For null
geodesics it then follows that \( \mathrm{d} t = \left( 1 - \phi \right) 
\mathrm{d} l \), where \( l \) is the Euclidean length. 
Integrating along to the light ray trajectory  and introducing the angular 
variables  \( \theta \) and \( \vartheta \) we find

\begin{equation}
  \delta \tau =\frac{ D_{OS} D_{OL} }{ D_{LS} } \left[ \frac{ 1 }{ 2 }
    \left( \theta - \vartheta \right)^2 - \frac{ 1 }{ 2 } \theta_E^2 \ln(
    \theta  / \theta_E )+ \alpha  \theta^2 \ln( \theta / \theta_E )
    \right].
\label{eq33}
\end{equation} 

  According to this, if \( \alpha = 0 \), we obtain the general
relativistic  time delay.  The extra term \( \alpha \, \theta^2 \, \ln(
\theta / \theta_E) \) results in a contribution for it when \( \theta \)
is greater than \( \theta_E \).  For the case of \( \theta = \theta_{E}
\), this contribution is null. However, the Einstein angle is not
a solution of the lens equation \eqref{eq28} even if \( \vartheta =
0  \), so this case is never achieved.  In other words,
equation~\eqref{eq33} means that the modified time delay is greater than
what is obtained in standard general relativity.

\section{Conclusion}
\label{conclusion}

  We have shown that all metric theories of gravity of the form
\( f(R) = R^n \)  produce gravitational waves propagating at the
velocity of light in vacuum.  In particular, the theory developed by
\citet{sobouti06} satisfies this condition.  We have also proved that
\citeauthor{sobouti06}'s theory produces an additional amount of lensing as
compared to standard general relativity calculations.

  Of course, more investigation on the physical and astrophysical side
of the theory developed by \citeauthor{sobouti06} needs to be done.
By no means can his theory be taken as a fundamental one, but rather as a
suitable candidate approximation at a certain scale.  Particularly,
more values of his \( \alpha \) parameter need to be calculated for
different astronomical environments.    Also, more development in the
theory of gravitational lensing needs to be done, in order to compare
directly with current astronomical data, particularly the anomalous
lensing observed in cosmology.

  As a final remark, we briefly mention that \citeauthor{sobouti06}'s theory
is not affected by the no--go theorem proposed by \citet{soussa03a} and
expanded by \citet{soussa03b}.  This is easy to see if we note that some of
the components of the perturbations made to the field
equations~\eqref{field-equations} do not necessarily scale with \( h^2 :=
h_{\mu\nu} \, h^{\mu\nu} \).  In fact, in the weak field approximation,
the time--component of equation~\eqref{field-equations} takes the form

\begin{equation}
  \frac{ 4 }{ 3 } \nabla^2 \phi - 2 \alpha \left[ - \frac{ 5 }{ 6 } +
    \frac{ \ln \left( 3 \alpha \right) }{ 3 } \right] \nabla^2 \phi +
    \frac{ 2 \alpha }{ 3 } \nabla^2 \phi \,  \ln \left(  -2 \nabla^2
    \phi \right) = \frac{  16 \pi G \rho }{ 3 }.
\label{eq50}
\end{equation}

\noindent From this equation and the value of \( \alpha \) given by
equation~\eqref{eq25b} it is clear that the gravitational
potential \( \phi \) may scale as \( \sqrt{ G M } \) without ever
reaching terms of the order of \( h^2 \).  This is the reason of why
\citeauthor{sobouti06}'s theory does not satisfy the conditions of the
no--go theorem and so, may account for an extra amount of lensing as
explained before.

  While this article was being refereed, and since its appearance in
the arXiv, a short comment was made by \citet{saffari07}.  This author
states that some of the calculations made by \citet{sobouti06} are wrong
and this is reflected in the circular velocity with a change of sign in
one of its terms.  In his conclusions he states that, although the main
results obtained by Sobouti are not strongly changed, they may affect
higher order corrections of the theory and in the limit applicable to
very compact objects.  Although our article does not reproduce Sobouti's
equations mentioned by Saffari, we were very well aware of the small
error made by Sobouti.  In fact, the corrections to Sobouti's calculations
have been published in the Thesis made by \citet{rosas06} (see equations
(4.17), (4.20)-(4.24) of Rosas--Guevara Thesis and compare them with
equations (2)-(8) in Saffari's article).  So, all the results discussed
in the present article are not affected in any manner by the small error
made by Sobouti, because the corrections were already included.

  Although Sobouti's theory seems very attractive it has a small
caveat which was discussed by him:  ``Actions are ordinarily
form invariant under the changes in sources.  Mass dependence of \( \alpha
\) (cf. equation~\eqref{eq25b}) destroys this feature and the claim for the
action--based theory should be qualified with such reservation in mind''.
While this fact is not enough to rule out the theory proposed by Sobouti,
one must not forget it.

  In summary, it all seems that \citeauthor{sobouti06}'s theory may play
the role of a good candidate  for a modified theory of gravity that can be
used in the understanding of different astrophysical phenomena, usually
described by dark matter.  The simplest way to do this in the future is
by direct comparisons with observed gravitational lenses (particularly
Einstein's rings) in clusters of galaxies.  We are working in this
direction and the results obtained will be published elsewhere.

\section{Acknowledgements}
  We thank X. Hernandez for the many discussions, suggestions and
comments made to the first draft of this article. We are also grateful
to J. Bekenstein for his comments regarding the bending angle and its
potential significance at cosmological scales.  We very much appreciate
the deep comments made by Y. Sobouti, particularly in writing down
explicitly the result to \( \text{O}(\alpha) \) of equation~\eqref{eq25},
i.e. relation~\eqref{eq25a}.  We also thank L.A. Torres for his comments
made to the first draft of the article.  We acknowledge the final reading
of the article, including its calculations, to T. Bernal. We thank an
anonymous referee for his comments which improved the final version of
the paper.  The authors gratefully acknowledge financial support from
DGAPA--UNAM~(IN119203) and (IN11307-3).  S.~Mendoza also acknowledges
financial support from CONACyT (41443).

\bibliographystyle{aa}
\bibliography{mod}

\end{document}